%input macro
%at the start of a tex file. Note that different printers may or
%may not automatically include an offset---if the text is off to
%one side adjust the commands \hoffset and \voffset by adding or
%removing a comment sign %.

\def\chaphead{}
\def\ni{\noindent}

\font\tfont=cmbxti10
\font\eightrm=cmr8
\font\eightit=cmti8
\font\sixrm=cmr6
\font\eightmit=cmmi8
\font\sixmit=cmmi6
\def\absmath{\textfont0=\eightrm \scriptfont0=\sixrm
	      \textfont1=\eightmit \scriptfont1=\sixmit}
\def\absfont{\let\rm=\eightrm \let\it=\eightit \rm\absmath}
\font\twelverm=cmr12
\font\twelveit=cmti12
\font\tenrm=cmr10
\font\twelvemit=cmmi12
\font\tenmit=cmmi10
\def\regmath{\textfont0=\twelverm \scriptfont0=\tenrm
	      \textfont1=\twelvemit \scriptfont1=\tenmit}
\def\peterfont{\let\rm=\twelverm \let\it=\twelveit \rm\regmath}
%
%\def\abstract#1{ {
%\centerline{\bf ABSTRACT}
%\vskip 12pt
%\let\rm=\eightrm \let\it=\eightit \rm\absmathpt
%{\parshape=1 0.5in 6.3truein \noindent #1
%
%  }\parshape=0}\vfill\eject}

%following macro defines vectors using arrows
  %This contradicts tex82 which has \b meaning
%underline instead of vector superscript. Skew is added to
%improve location of arrow.
%following macro defines vectors as boldface italic. Note in this case that
%all characters regarded as being in the argument of the macro will be
%boldface.
\newfam\vecfam

\textfont\vecfam=\tfont \scriptfont\vecfam=\seveni
\scriptscriptfont\vecfam=\fivei

 %error function

\def\spose#1{\hbox to 0pt{#1\hss}}

\font\eightrm=cmr8

\def\s{\ifmmode \widetilde \else \~\fi} %produces tilde in mathmode or
%horizontal mode.
     
%\def\={\overline}
\def\section{\S}
\newcount\notenumber
\notenumber=1
\newcount\eqnumber
\eqnumber=1
\newcount\fignumber
\fignumber=1
\newbox\abstr

%\numberpara produces numbered paragraphs with extra space and no indentation

\def\s{{\rm\,s}}

%\note macro produces sequentially numbered footnotes at bottom of page
%\foot macro produces sequentially numbered footnotes inserted in text
\def\note#1{\footnote{$^{\the\notenumber}$}{#1}\global\advance\notenumber by 1}
\def\foot#1{\raise3pt\hbox{\eightrm \the\notenumber}
     \hfil\par\vskip3pt\hrule\vskip6pt
     \noindent\raise3pt\hbox{\eightrm \the\notenumber}
     #1\par\vskip6pt\hrule\vskip3pt\noindent\global\advance\notenumber by 1}

%\abstract macro makes abstracts
\def\abstract#1{\setbox\abstr=\vbox{\hsize 5.0truein{\par\noindent#1}}
    \centerline{ABSTRACT} \vskip12pt \hbox to \hsize{\hfill\box\abstr\hfill}}
     
%\Dt and \dt put Newton's notation dots above upper and lower case chars
\def\Dt{\spose{\raise 1.5ex\hbox{\hskip3pt$\mathchar"201$}}}    % upper case
\def\dt{\spose{\raise 1.0ex\hbox{\hskip2pt$\mathchar"201$}}}    % lower case

% equation numbering
%\new macro produces sequentially numbered equations by writing \eqno(\new)
%at end of displayed equations
\def\new{{\rm\chaphead\the\eqnumber}\global\advance\eqnumber by 1}
%to refer to an equation which is 5 equations back, write "equation (\ref5)"
\def\ref#1{\advance\eqnumber by -#1 \chaphead\the\eqnumber
     \advance\eqnumber by #1 }
%\last macro is like \new except counter is not advanced. Useful for equations
%which are in parts a and b.
\def\last{\advance\eqnumber by -1 {\rm\chaphead\the\eqnumber}\advance
     \eqnumber by 1}
%to name an equation, place command "\eqnam{\Poisson}" before equation, and
%thereafter "equation(\Poisson)" will generate the proper equation number.
\def\eqnam#1{\xdef#1{\chaphead\the\eqnumber}}
     
%figure numbering
%\nfig macro assigns number to a figure
\def\nfig{\chaphead\the\fignumber\global\advance\fignumber by 1}
%\nfiga permits a,b,c etc. to be added to figure number
\def\nfiga#1{\chaphead\the\fignumber{#1}\global\advance\fignumber by 1}
\def\rfig#1{\advance\fignumber by -#1 \chaphead\the\fignumber
     \advance\fignumber by #1}
\def\fignam#1{\xdef#1{\chaphead\the\fignumber}}
%reference macros. To generate reference to a paper in Ap.J. volume 300, p.123
%write \apj{Claus, S. 1990.}{300}{123}

%\lta and \gta produce > and < signs with twiddle underneath
\def\lta{\mathrel{\spose{\lower 3pt\hbox{$\mathchar"218$}}
     \raise 2.0pt\hbox{$\mathchar"13C$}}}
\def\gta{\mathrel{\spose{\lower 3pt\hbox{$\mathchar"218$}}
     \raise 2.0pt\hbox{$\mathchar"13E$}}}
     
%\sec produces arcsec symbol so that 3\sec5 produces 3."5 with the second
%symbol and the period aligned.

\magnification=\magstep1
%\hoffset=1.0truein
%\voffset=0.8truein
\parskip=3pt

\def\ni{\noindent}
\def\refind{\noindent \hangindent=2pc \hangafter=1}

\baselineskip= 12pt

\centerline{\bf Asymmetries of Solar p-mode Line Profiles}
\vskip 0.7truecm
\centerline{Douglas Abrams and Pawan Kumar\footnote{$^1$}
{Alfred P. Sloan Fellow \& NSF Young Investigator}}
\bigskip
\centerline{Department of Physics}
\medskip
\centerline{Massachusetts Institute of Technology, Cambridge, MA 02139}
\bigskip\bigskip
\centerline{\bf Abstract}
\bigskip

Recent observations indicate that solar p-mode line profiles are not 
exactly Lorentzian, but rather exhibit varying amounts of asymmetry about 
their respective peaks. We analyze p-mode line asymmetry using both a 
simplified one-dimensional model and a more realistic solar model.  We
find that the amount of asymmetry exhibited by a given mode
depends on the location of the sources exciting the mode, 
the mode frequency, and weakly on the mode spherical harmonic degree, 
but not on the particular mechanism or location of the damping. 
We calculate the dependence of line asymmetry on source location 
for solar p-modes, and provide physical explanations of our 
results in terms of the simplified model. A comparison of our 
results to the observations of line asymmetry in velocity spectra 
reported by Duvall et al. (1993) 
for modes of frequency $\sim$ 2.3 mHz suggests that the 
sources for these modes are located more than 325 km beneath the photosphere.
This source depth is greater than that found by Kumar (1994) for acoustic 
waves of frequency $\sim$ 6 mHz. The difference may indicate that 
waves of different frequencies are excited at different depths in the 
convection zone.  
We find that line asymmetry causes the frequency obtained from a Lorentzian 
fit to a peak in the power spectrum to differ from the corresponding 
eigenfrequency by an amount proportional to a dimensionless asymmetry parameter 
and to the mode linewidth.

\bigskip
\ni{\it Subject headings:} Sun --- oscillations: sun --- p-modes

\vfill\eject
\centerline{\bf 1. Introduction}
\medskip

Most previous work in helioseismology has relied on the 
assumption that solar p-mode line profiles are Lorentzian, 
which is expected for the power spectrum of 
randomly forced, damped harmonic oscillators.
However, recent results indicate that this may be 
only approximately true.  Duvall et al. (1993) 
have presented observational evidence from their South Pole 
data that p-mode line shapes are not exactly Lorentzian, but rather exhibit 
varying amounts of asymmetry about their respective peaks.  
This is a difficult observation which has not yet been reproduced by 
other groups.  However, several authors have investigated 
the problem theoretically, and have found that line profiles are 
asymmetrical whenever a localized excitation source is present 
(Gabriel, 1992, 1993, and 1995; Roxbourgh \& Vorontsov, 1995; see also Figure 2 of 
Kumar et al., 1989).  Therefore, we believe that the line asymmetry observations of Duvall et al. will be validated when more 
accurate measurements are available.

The excitation of p-modes is thought to occur in a thin layer near the top of 
the solar convection zone.  All of the previous work on line asymmetry 
indicates that the degree of asymmetry depends on the depth of 
the excitation sources.  Therefore, p-mode line shapes may provide 
a means of locating the acoustic sources responsible for exciting 
modes of various frequencies.

In this paper, we present a detailed analysis of p-mode line profiles for 
both a simplified one-dimensional model and a realistic solar model, 
previously used by Kumar et al. (1994) to study peaks in the 
power spectrum above the acoustic cutoff frequency.  
For the one dimensional model problem, we analyze the 
dependence of line asymmetry on source depth, mode dissipation (including 
linewidth and location of damping), mode frequency, and $\ell$; we also 
investigate the error introduced in p-mode frequency determination by 
Lorentzian fits to the power spectrum.  Our results for the one-dimensional 
problem are contained in section 2.  For the realistic solar model, we investigate
the dependence of line asymmetry on source depth, mode frequency and $\ell$, 
and  also examine errors in eigenfrequency determination.  These results are 
contained in section 3.  We summarize our main conclusions in section 4.

\bigskip
\centerline{\bf 2. Model problem}
\medskip

The following (homogeneous) one-dimensional wave
equation describes adiabatic solar oscillations in the Cowling approximation 
(Deubner and Gough, 1984):
$$ {d^2\psi_\omega\over d r^2} + \left[{\omega^2\over c^2} 
- V(r)\right]\psi_\omega = 0, \eqno(1) $$
where $\psi_\omega = \rho^{1/2} c^2  {\rm div} \vec {\xi_\omega}$, $\vec {\xi_\omega}$ is a 
Fourier component of the fluid displacement, $\rho$ is 
the equilibrium density, and $c$ is the sound speed.  
The effective acoustic potential is given by $V(r) \approx 
{\ell(\ell + 1) / r^2} + {\omega_{ac}^2 / c^2}$, 
where $\omega_{ac}$ 
is the acoustic cutoff frequency.  The first term in the potential 
determines a mode's lower turning point, while the 
second term peaks near the temperature minimum and causes acoustic 
waves to be reflected.  Solar p-modes can be modelled reasonably well 
by setting the sound-speed equal to one everywhere and 
using the following simple form for the potential 
(see also Kumar and Lu, 1991; Kumar et al., 1994):  
$$ V(r)=\cases{\infty&  for $r\le 0$,\cr  0& for $0<r<a$,\cr 
\alpha^2 \ {\rm (constant)} & for $r \ge a$,\cr} \eqno(2) $$
where $a$ is the sound 
travel time from the lower to the upper turning point of a given mode 
and $\alpha$ is the acoustic cutoff frequency at the temperature minimum. 
Waves of angular frequency less than $\alpha$ are trapped below $a$, while 
waves of angular frequency greater than $\alpha$ can propagate to 
infinity; thus, for frequencies less than $\alpha$ there is a discrete 
spectrum of real eigenfrequencies. Neglecting a weak 
dependence on mode frequency, different values of $a$ correspond 
to modes of different degree $\ell$.

We now add a damping term and a source term to equation (1).  
Although the damping processes affecting solar p-modes are 
quite complex, we assume that for the present purposes 
they may be modeled by a viscous damping force.  
Upon adding these two terms, we obtain
$${d^2\psi_\omega\over dr^2} + i \omega \gamma_\omega(r) 
\psi_\omega + \left[ \omega^2 - V(r) \right] 
\psi_\omega = f_\omega(r), \eqno(3) $$
where $f_\omega (r)$ and $\gamma_\omega (r)$ are the Fourier components 
of the source function and of the coefficient of the damping term, respectively.  
As long as the region of excitation is much smaller than the wavelengths 
considered, we may use $f_{\omega}(r) =
 S_{\omega} \delta(r-r_s)$.  Since the source power spectrum 
$S_{\omega}$ varies negligibly over a typical mode linewidth, it does not affect our 
calculations of asymmetry and we set it equal to unity for all frequencies.  We 
consider two forms for $\gamma_{\omega}(r)$.  One case we consider is 
$\gamma_{\omega}(r) = \Gamma_{\omega}$ (independent of $r$), where 
$\Gamma_{\omega}$ is the (frequency-dependent) linewidth.  
The other is $\gamma_{\omega}(r) = \Gamma_\omega' \delta(r-r_d)$, 
which is a better approximation to solar p-mode dissipation.  
These two extreme cases should demonstrate whether or not line asymmetry
has any dependence on the details of the damping process.

Equation (3) is easily solved to obtain the power spectrum seen by an 
observer in the photosphere, and individual peaks in spectra generated in this 
manner do indeed exhibit varying degrees of asymmetry (see Figure 1).  
In the next two 
subsections, we examine the asymmetry when the source is inside or outside 
the well.  In \S2c 
we examine the effects of asymmetry on the accuracy of determining the 
system's eigenfrequencies using Lorentzian fits to the peaks.

Before proceeding, we introduce a method of quantifying 
line asymmetry which will be used for the rest of this paper.  
We decompose the 
observed power spectrum in the neighborhood of a peak corresponding to a 
mode $\alpha$ into even and odd functions.  Since the odd function is zero 
at the peak and again far from the peak, its magnitude has a maximum at 
some intermediate distance from the peak, typically less than one 
linewidth.  The ratio of the maximum magnitude of the odd function to the 
maximum magnitude of the even function is a dimensionless measure of the 
asymmetry which we denote by $\eta_\alpha/100$. Then $\eta_\alpha$ is
the percentage line asymmetry of mode $\alpha$. The sign of $\eta_\alpha$
is taken to be positive or negative according to whether there is more 
power on the high- or low-frequency side of the peak, respectively.

\bigskip
\ni{\bf 2a. Source inside the well}
\medskip

	When the source is inside the well and damping is uniform (i.e., 
$\gamma_{\omega}(r) = \Gamma_{\omega}$ everywhere), the solution to 
equation (3) is given by
$$\psi_\omega(r)= -\left({\sin k r_s\over k\cos k a +
k_1 \sin k a}\right) e^{- k_1 (r-a)}, \eqno(4)$$
where $k = \sqrt{\omega^2 + i\omega\Gamma_{\omega}}$ and $k_1 = 
\sqrt {\alpha^2 - \omega^2 - i\omega\Gamma_\omega}$.  When 
$\Gamma_{\omega}\ll\omega$ (as is the case for solar p-modes) and $\omega \approx \omega_\alpha$, an eigenfrequency, the velocity power spectrum 
seen by an observer at location $r$ is given by
$$ P_\omega (r) \equiv \omega^2|\psi_\omega(r)|^2 
\approx{\omega^2 e^{- 2 k_1 (r-a)}\over(a\alpha)^2}
\left[{\sin^2 r_s \omega \over (\Delta\omega)^2 + 
\Gamma^2_\omega / 4} \right], \eqno(5)$$
where $\Delta\omega=\omega-\omega_\alpha$.  The line-profile depends on the 
source position through the factor $\sin^2r_s\omega$ in the numerator.  When 
$\sin r_s\omega_\alpha \approx 1$, the variation of the numerator with frequency 
over the mode linewidth is small, and the line profile is nearly Lorentzian.  
(In all cases, the variation over a linewidth of the factor appearing outside the 
square brackets is negligible.)  However, when $\sin r_s\omega_\alpha \approx 
0$, the peak is quite asymmetrical. Note that the most asymmetrical peaks, 
when the source lies inside the well, 
therefore correspond to modes which are excited to small amplitudes.

For the discussion that follows, it will be useful to rewrite the
 power spectrum in the neighborhood of a mode frequency as:
$$ P_\omega (r) \approx C \left[ {\sin^2(\delta + r_s\Delta\omega) 
\over (\Delta\omega)^2 + \Gamma^2_\omega/4}\right], \eqno(6)$$
where $C$ is approximately constant, $\delta = \omega_\alpha(r_s-r_0)$, and 
$r_0$ is the closest number to $r_s$ satisfying $\sin r_0 \omega_\alpha = 0$. 
(Note that $r_0$ may be greater than $a$; in particular, it need not 
correspond to a node of the eigenfunction in question.)  
It is apparent from equation (6) and the discussion above that the asymmetry 
parameter $\eta_\alpha$ has the same sign as $\delta$, and that the magnitude of 
$\eta_\alpha$ is maximized ({\it i.e.} peaks are very asymmetrical) when $\delta$ is small.  
When $\delta=0$, however, the source lies exactly at a node, 
and it is meaningless to speak of line asymmetry, 
since the Lorentzian peak is completely suppressed.

We have checked the dependence of line asymmetry on source 
location, mode frequency, linewidth, and $\ell$, when the sources lie
within 1000 km (100 seconds of sound travel time) of the upper turning point 
(see Figure 2).  Moving the sources deeper causes $\eta_\alpha$ 
to become more positive, except when the source passes through a node, 
in which case $\eta_\alpha$ jumps 
discontinuously from a large positive to a large negative value.
This follows from the dependence of $\delta$ on $r_s$: as the source
is moved deeper into the well, $\delta$ decreases monotonically except
when $\eta_\alpha$ passes through zero, in which case 
$\delta$ jumps discontinuously from $-\pi/2$ to $\pi/2$.

The behavior of the asymmetry as a function of mode frequency depends on the source location: 
for source locations within 400 km of the upper turning point, the magnitude of $\eta_\alpha$ 
is greatest for low-frequency modes, while for some deeper source locations 
(for example, 800 km depth), it is greatest for high-frequency modes.  
(Since the asymmetry for a given mode depends on the locations of its eigenfunction's nodes 
relative to the source, the modes with the most asymmetrical power spectra continue to change 
as the source is moved deeper still.)
  
Increasing the linewidth with other parameters fixed 
increases the magnitude of the asymmetry, since the variation of the
numerator in equation 6 over the extent of the peak is effectively 
increased.  Finally, with the source restricted to 
lie near the top of the cavity, the magnitude of the asymmetry 
increases with increasing cavity length (decreasing $\ell$ value), 
since the numerator in Equation 6 varies more rapidly.

	These dependences of the asymmetry on the linewidth 
and effective cavity length are quite general and do not depend on 
the detailed nature of the potential.

	We have also considered the case when the 
damping is localized as a delta function 0-250 km 
below the upper turning point; we find the dependence of
lineshape on the nature and location of damping to be very
weak.  In order to make meaningful 
comparisons between the cases of global and local damping, 
we match the imaginary parts of the respective eigenfrequencies.  
For cavity lengths corresponding to low degrees ($\ell \approx 5$), we 
consider linewidths of up to 15 $\mu$Hz, while for cavity lengths 
corresponding to higher degrees, we consider larger linewidths; for $\ell \gta 350$, 
we consider linewidths of up to 50 $\mu$Hz.  In this parameter regime, 
we find that if $1\% \lta |\eta_\alpha| \lta 10\%$, 
changing the damping location changes $\eta_\alpha$ 
by less than 10\% of its total value; we also find that results 
are similar to those obtained using uniform damping \footnote{$^1$} 
{Other damping schemes yield similar results as well; 
in particular, when the damping 
is non-zero above a certain depth (near the upper turning point), 
or when it is nonzero only in a finite range at the top of the well, 
the behavior of the asymmetry is essentially unchanged.}.

Thus, when the source lies just beneath the upper turning point, 
lineshapes depend strongly on source location, mode frequency, 
linewidth, and $\ell$, but only weakly on the type and location 
of damping.  Placing the source in the evanescent region leads 
to similar conclusions; that case is discussed next.

\medskip
\ni{\bf 2b. Source outside the well}
\medskip

When damping is uniform throughout the well, the solution of equation (3)
for the amplitude seen by an observer at location $r$, due to a point 
source at $r_s$ ($r>r_s>a$), is given by:

$$\psi_\omega(r)= \left[{2k \cos k a \sinh k_1 (r_s -a ) + 2k_1 \sin k a \cosh k_1 (r_s-a)
\over k\cos k a + k_1 \sin k a}\right]{e^{- k_1 (r-a)}\over 2k_1}, \eqno(7)$$
where $k$ and $k_1$ are defined as before.  
However, we gain more insight in this case 
by considering separately the contributions to the total amplitude 
due to waves which travel along different paths from the source to the
observer. (This is the approach taken by Duvall et al., 1993.)
In particular, the total observed amplitude is given by the sum of three parts: 
(1) the wave which 
travels directly outward from the source to the observer 
(hereafter the direct wave), (2) the wave which first travels toward
the cavity and is reflected back toward the observer at the potential step 
(hereafter the reflected wave), and (3) the sum of the infinite sequence of 
waves which arises due to multiple reflections in the cavity (hereafter the 
cavity wave).
	
	The total amplitude seen by an observer is thus given by:
$$ \psi_{\omega , tot}(r) = \psi _{\omega , dir}(r) + 
\psi _{\omega , ref}(r) + \psi _{\omega , cav}(r) =  
{- e^{-k_1 (r-r_s)} \over 2 k_1} $$
$$ - {e^{-k_1 (r + r_s - 2a)} \over  2k_1} 
\left({k_1+ik \over k_1-ik}\right) + 
{e^{-k_1 (r + r_s - 2a) + 2ika} k 
\over 2kk_1 +i(k_1^2 - k^2)} 
\cdot \left [1 + e^{2ika} \left({ik+k_1 
\over ik - k_1}\right)\right] ^ {-1}. \eqno(8)$$
The first two terms are roughly
constant in magnitude and phase over a typical linewidth (up to 50 $\mu {\rm 
Hz}$), while the third term (the cavity wave) exhibits resonance behavior.  In 
particular, constructive interference occurs when the waves 
due to multiple reflections in 
the cavity arrive at the observer in phase; the frequencies at which this 
occurs are the eigenfrequencies.  Therefore, for all cases of interest, the amplitude of the cavity wave is much greater than the amplitudes of the direct and reflected waves at an eigenfrequency.  The magnitude of the 
cavity wave in the neighborhood of a mode eigenfrequency 
varies symmetrically about the peak.  However, interference 
between the cavity wave and the sum of the direct and reflected waves causes 
the total power seen by an observer to be asymmetric about the peak.
	
	For small linewidths (specifically, in the limit that 
$k$ and $k_1$ are real), the cavity wave leads the direct wave 
by $\pi/2$ in phase when $\omega$ is equal
to an eigenfrequency; this phase difference is due to the phase 
shifts experienced by a wave upon entering or leaving the well, 
plus the phase shift due to travel from the top 
of the cavity to the bottom and back, with an inverting reflection at the bottom.
Since the reflected wave is smaller in magnitude
than the direct wave, it follows that at resonance, the cavity wave always 
leads the sum of the direct and reflected waves in phase.  Furthermore, the 
phase of the cavity wave is a monotonically increasing function of frequency.  
Thus the observed asymmetry is always negative, while the magnitude 
of asymmetry depends on the differences in 
amplitude and phase between the three interfering terms.  
(It suffices to evaluate these at the source location, since the three waves 
experience the same phase change and attenuation in traveling from the source 
to the observer.)  The relative amplitudes 
and phases of the three waves near a mode frequency are shown in Figure 3.  These 
conclusions are unaltered by the presence of damping.

	We have investigated the dependence of the asymmetry on 
source location, mode frequency, mode degree, and linewidth (see Figure 2).  
Moving the source away from the cavity leads to more asymmetrical line profiles 
because the amplitude of the cavity wave (evaluated at the source location) 
decreases due to attenuation in the evanescent zone, 
while the amplitude of the direct wave is unaffected; 
the interference of the direct wave with the cavity 
wave therefore has a more pronounced effect.  Increasing mode frequency 
while keeping the linewidth fixed, on the other hand, 
leads to less asymmetrical profiles because the amplitude of the cavity wave at 
resonance $increases$ with increasing mode frequency, relative to the sum of 
the direct and reflected waves.  We find that increasing the 
cavity length (decreasing $\ell$) while keeping the linewidth fixed 
yields more asymmetrical profiles.  
Finally, increasing the linewidth yields more asymmetrical line profiles, 
since increasing the damping decreases 
the amplitude of the cavity wave without significantly 
affecting the amplitudes of the direct or reflected waves.

As in the case when the source is inside the well, line profiles
depend very weakly on the nature and location of damping: the numbers 
quoted in this respect in \S2a apply here as well.

We have shown that line asymmetry, when the source is above or 
below the upper turning point, depends strongly on source location, 
linewidth, mode frequency, and $\ell$, but 
only weakly on the nature and location of damping.  
Line asymmetry also leads to errors in determining 
the system's eigenfrequencies using Lorentzian fits to the peaks; that 
is discussed next.

\medskip
\ni{\bf 2c. Errors in eigenfrequency determination}
\medskip

Errors in eigenfrequency determination by Lorentzian fits to 
observed power spectra occur for two reasons.  First of all, the 
observed peak in a given line profile may be shifted from the corresponding 
eigenfrequency.  Secondly, the frequency of best fit may differ from  
the frequency corresponding to the observed peak.  
We have tested eigenfrequency determination error in the 
model problem by fitting Lorentzians to profiles generated for 
different source locations, mode frequencies, cavity lengths, and 
linewidths, including cases when the source is inside or outside the 
well and when the damping is global or local.  

We find a simple relationship between percent asymmetry and the 
amount $\delta\nu_\alpha$ by which the frequency obtained from a 
Lorentzian fit to the power spectrum of a mode $\alpha$ differs from the 
corresponding eigenfrequency.  Expressed as a percentage of the 
corresponding linewidth ($\Gamma_\alpha$), $\delta\nu_\alpha$ is 
proportional to the percentage asymmetry, $\eta_\alpha$: 
$$ \delta \nu_\alpha = \left[{b\over 100}\right] 
\Gamma_\alpha \eta_\alpha, \eqno(9) $$ 
where $b$, the constant of proportanality, depends only on the type of 
damping used, and not on mode frequency, linewidth, $\ell$ value, or source 
location.  For spectra generated with global damping, $b \approx 1.6$, 
whereas for spectra generated with local damping, $b \approx 1.1$.  
(See Figure 5 of \S 3c).

\bigskip
\centerline{\bf 3. Line asymmetry for p-modes of the Sun}
\medskip

We have also analyzed line asymmetry using a solar model due to 
Christensen-Daalsgard (1991). Our calculations of p-mode power spectra 
include radiative damping and stochastic mode excitation due to 
turbulent convection, but ignore dissipation due 
to the interaction of modes with turbulent convection.  
Based on the calculations described in \S2, 
we do not expect our results to depend on the type of damping used; 
however, due to the fact that the linewidths we 
calculate are smaller than the observed linewidths, 
we must interpret our numerical results taking into account 
the dependence of asymmetry on linewidth 
found for the simple model of \S2.

Theoretical power spectra are computed by solving 
for Green functions of the nonadiabatic oscillation 
equations in the Cowling approximation.  The details 
of this calculation are given by Kumar (1994).  
As described therein, the mean velocity power spectrum 
due to stochastic excitation by turbulent convection is given by:
$$\left<P_\omega \right> \approx {\omega^2 \over {R_\odot}^2} 
\int dr \int dr_0 \left|\rho(r_0) {d\over dr_0} G_\omega (r;r_0)\right|^2 
S_\omega(r_0), \eqno(10)$$
where $G_\omega(r;r_0)$ is the Green function for a source term 
in the momentum equation, and $S_\omega(r_0)$ is the source strength, 
which we consider to be the Reynold's stress.  
The outer integral is over the region in the photosphere where the 
optical line is formed, and the inner integral is over the source region.
The p-mode power spectrum calculated using the above equation 
depends on the derivative of the Green function because we consider 
only quadrupole sources of sound waves, which correspond to the 
derivative of the Reynold's stress; this derivative is transferred to the 
Green function after integration by parts.  The power spectrum for 
dipole sources, obtained using equation (10) with $G_\omega$ 
in place of $dG_\omega/dr$, yields different lineshapes.

The source strength for quadrupole acoustic emission is given by Kumar (1994):
$$ S_\omega(r_0) \sim {H^4 v_H^3 \over 1 + (\tau_H \omega)^{7.5}}, \eqno(11)$$
where $H$, $\tau_H$, and $v_H$ are, respectively, the correlation lengths 
(``mixing lengths'') , correlation times, and velocities of energy bearing
convective eddies at radius $r_0$ ($v_H \approx H / \tau_H$).  
Due to its strong dependence 
on convective velocity, the source strength is expected to be a 
sharply peaked function of position (see Figure 1 of Kumar, 1994).  
We treat the unknown source strength as a Gaussian of width 
50 km (this is roughly what is expected if we take the convective velocity
as given by standard mixing length theory), 
and vary its peak position within 
the top 1200 km of the convection zone.  
We compute the p-mode line profiles and asymmetries which, 
when compared with observations, should yield the radial
location of the acoustic sources.  
We note that the equilibrium solar model was
calculated using standard mixing length theory.

In the next two subsections,  we present the results of our solar model 
calculations for the dependence of line asymmetry on source location, mode 
frequency, and mode degree, and explain our results in terms of the 
simplified model.  In \S3c we discuss errors in eigenfrequency 
determination introduced by line asymmetry.

\medskip
\ni{\bf 3a. Solar model results}
\medskip

We have calculated solar p-mode power spectra for $\ell$ between 5 and 500, 
frequencies in the range $\sim 1-5$ mHz, and source locations in the top 
1200 km of the convection zone.  Peaks exhibit varying amounts of asymmetry,
depending on source location, mode frequency, and mode degree (see Figure 4).

Moving the sources deeper causes the asymmetry parameter 
$\eta_\alpha$ to become more positive, 
unless the source passes through a node, in which case $\eta_\alpha$ 
jumps from a large positive to a large negative value.  
(There is some deviation from this pattern when the sources 
lie in the top 30 km of the convection zone.)

When the source depth is less than 400 km, the most asymmetrical peaks 
(with $|\eta_\alpha| \approx 10\%$) correspond to 
low-frequency modes, while peaks corresponding to modes 
above 3 mHz show very little asymmetry ($|\eta_\alpha| < 4\%$).  
The asymmetry for mode frequencies below 3 mHz is negative, 
{\it i.e.} there is more power on the lower frequency side of the peak.
In contrast, for some deeper source locations 
(for example, depths between 800 and 1000 km), 
the most asymmetrical peaks correspond to high-frequency modes, while 
peaks corresponding to modes below 3 mHz show little asymmetry.  
In this case, the asymmetry of the most asymmetrical peaks is positive.
In general, which modes have the most asymmetrical power spectra, 
and the sign of the corresponding asymmetries, depends on the location
of the sources, and changes as the sources are moved deeper still.

For any source location and mode frequency, 
the magnitude of the asymmetry is a weakly increasing function 
of $\ell$; $\eta_\alpha$ changes by no more than a few percentage 
points over the range $\ell = 5 - 500$.

Duvall et al. (1993) found negative asymmetry in p-mode velocity 
spectra corresponding to modes of frequency 2.2-2.5 mHz and degrees in 
the range 157-221; from their data 
we find that $\eta_\alpha = -2.5\%$ for these modes.  
Allowing an uncertainty in $\eta_\alpha$ of ${^+/_-}$ 0.5\%, 
our numerical calculations reproduce this result provided 
we assume that the sources for modes of this frequency 
lie at a depth of 325-525 km beneath the photosphere.  
However, our calculated linewidths are less than the observed linewidths 
in this frequency range, which implies that for any source location, 
we underestimate the magnitude of $\eta_\alpha$.  
Since $|\eta_\alpha|$ for these modes 
is a decreasing function of source depth 
for source locations within the top few scale heights of the 
convection zone, this technique therefore yields a lower bound 
on the source depth for these modes of 325 km.

Duvall et al. found {\it positive} line asymmetry in the 
intensity power spectra of the same modes.  This is a puzzling result.  
At any given location in the solar atmosphere, the dynamical 
oscillation equations may be solved for the pressure and 
density perturbations ($\delta p$ and $\delta \rho$) 
in terms of the radial displacement function ($\xi_r$), 
the radial wave number, and equilibrium properties of the atmosphere.  
In the model problem of \S2, it can be shown analytically that 
$\xi_r$ is the only one of these quantities which varies appreciably 
over a mode linewidth, and numerical calculations 
show this to be the case for the solar model.  As a result we find that the
perturbation to any thermodynamic quantity is essentially proportional 
to the radial displacement function. Any reversal of asymmetry must 
therefore be due to a subtle detail of the process by which the observed 
flux is modulated by pulsation.

\medskip
\ni{\bf 3b. Explanation of solar model results}
\medskip

The behavior described above is in qualitative agreement with the 
results obtained using the simplified model of \S2, except for the 
dependence of asymmetry on source position when the sources 
lie in the top 30 km of the convection zone.  
This disagreement is most likely due to the more 
complicated behavior of the solar potential at the top of the convection zone.
However, for the extremely limited range of source 
locations involved, the variation in asymmetry due to different source locations
is correspondingly small - less than 15\% of the total asymmetry.  

In the simple model problem, the asymmetry depends strongly on the 
cavity length (the effective mode degree), 
while in the solar model the asymmetry is a very weak function of degree.  
The difference is accounted for by the increase of linewidth with degree 
in the solar model, which cancels the effect on the asymmetry of 
decreasing the effective cavity length.  
(The dependence of linewidth on $\ell$ calculated using the solar model 
is similar to the observed dependence; therefore, we do not expect 
this result to be affected significantly by the discrepancies between 
calculated and observed linewidths mentioned earlier.)  
 
As mentioned previously, the derivative used in calculating power 
spectra from the solar model has an important effect on the shapes of 
lines in the spectra (see equation 10).  In calculating power spectra 
using the simple model of \S2, no derivative is explicitly used.  
However, the dependent 
variable used there is $\psi = \rho^{1/2}c^2 {\rm div} \vec{\xi}$, 
which is approximately equal to $\rho^{1/2}c^2 d\xi_r/dr$ 
near the surface.  Due to the derivative present in the definition of $\psi$,
the shapes of lines in spectra calculated using the simple model agree 
qualitatively with the shapes of lines in spectra calculated using the solar 
model. (Note that the choice of variables determines the form of the 
acoustic potential. For the set of variables used here, the effective 
potential for p-modes may be approximated by the square-well potential 
of equation 2; for other sets of variables not involving derivatives - for 
example, the set used by Gabriel, 1992, and 
Roxbourgh \& Vorontsov, 1995 - the acoustic potential bears little 
resemblance to the square-well potential.)

\medskip
\centerline{\bf 3c.  Errors in eigenfrequency determination}
\medskip

As in the one-dimensional model problem, we find that for a mode 
$\alpha$, the frequency shift $\delta\nu_\alpha$ (see \S2c), expressed as a 
percentage of the linewidth $\Gamma_\alpha$, is proportional to the 
percentage asymmetry, $\eta_\alpha$.  (See Figure 5).  The constant of 
proportionality $b$, defined by equation (9), is found to be 1.5 in the solar 
case.  The linear relationship holds regardless of source location, mode 
frequency, or mode degree, as in the one-dimensional problem. 

\bigskip
\centerline{\bf 4. Summary and discussion}
\medskip

The shapes of lines in p-mode power spectra are, in general, 
not symmetric about the peak.  We find that the magnitude and sense of 
asymmetry depend strongly on the location of the sources and on mode 
frequency, and weakly on mode degree ($\ell$).

We define a dimensionless measure of 
asymmetry for a given mode by decomposing its power spectrum in the 
neighborhood of the peak into even and odd components; the 
ratio of the peak amplitudes of the two components, expressed as a percentage,
measures the magnitude of the asymmetry.  We define the asymmetry to be 
positive (or negative) when there is more power on the high-frequency 
(or low-frequency) side of the peak.
  
We have calculated the line asymmetry for p-modes of a simplified 
one-dimensional 
problem, as well as a solar model, for various source locations.  Line 
asymmetry is found to have similar behavior in both cases. 
In particular, we find that moving the sources deeper causes the 
asymmetry to become more positive, unless the source passes 
through a node of the mode eigenfunction, in which case the asymmetry 
changes discontinuously from a large positive to a large negative value.  
The magnitude of the asymmetry increases weakly with mode degree: 
for fixed source location and approximately constant mode frequency, 
the absolute value of the asymmetry parameter increases 
by no more than a few percentage points over the range 
$\ell =$ 5 to 500.  For the one-dimensional problem, the asymmetry 
has no significant dependence on the location or nature of damping as
long as mode linewidth is held fixed.

Physically, line asymmetry may be understood in terms of a 
wave-interference model, in which waves travel along different 
paths from the source to the observer, as suggested by 
Duvall et al. (1993).  However, we find 
that the line asymmetry does not depend significantly on phase shifts due 
to nonadiabatic effects in evanescent regions, as Duvall 
et al. suggested.

By comparing the asymmetries of theoretically calculated profiles to 
the asymmetries of the profiles observed by Duvall et al. (1993) 
for modes of frequency $\sim$ 2.3 mHz, we set a lower bound on the 
source depth for these modes of 325 km (measured with respect 
to the bottom of the photosphere).  
This is greater than the depth of 140 ${^+/_-}$ 60 km 
found by Kumar (1994) for the sources exciting acoustic waves
above the acoustic cutoff frequency ($\sim$ 5.3 mHz).  
Based on the theory of wave generation by turbulent convection and 
standard mixing length theory, we expect the emission of acoustic 
waves to occur deeper in the convection zone for waves of 
lower frequency.  The difference between the present results and 
the results of Kumar (1994) may be a confirmation of this prediction.

We find that line asymmetry causes the frequency obtained from a 
Lorentzian fit to a given peak in the power spectrum to differ from the 
corresponding eigenfrequency by an amount proportional to the mode 
linewidth and to the asymmetry of the peak.  
This holds for both the square-well potential model and the more realistic 
solar model.  As a percentage of the linewidth, this frequency shift for the 
solar model is about 1.5 times the percent asymmetry of the peak.

Duvall et al. reported that the sense of line asymmetry is different in 
velocity and intensity power spectra; in particular, for modes of frequency 
below 3 mHz, they found negative asymmetry in velocity spectra and positive 
asymmetry in intensity spectra. This is a puzzling result for which we
have no explanation.

\bigskip
\ni{\bf Acknowledgements:}
We thank Tom Duvall for sending us the data used in \S3a, 
and Eliot Quataert for many useful comments and discussions.
This work was supported by NASA grant NAGW-3936.

\vfill\eject
\centerline{\bf REFERENCES}
\bigskip

\refind Duvall, T.L.Jr., Jefferies, S.M., Harvey, J.W.,  Osaki, Y.,
and Pomerantz, M.A. 1993, ApJ, 410, 829

\ni Gabriel, M. 1992, AA, 265, 771

\ni Gabriel, M. 1993, AA, 274, 935

\ni Gabriel, M. 1995, AA, 299, 245

\ni Goldreich, P. and Keeley, D.K. 1977, ApJ, 212, 243

\ni Goldreich, P. \& Kumar, P. 1990, ApJ, 363, 694

\ni Goldreich, P., Murray, N., \& Kumar, P. 1993, ApJ, 424, 466

\ni Kumar, P., \& Lu, E. 1991, ApJ, 375, L35

\ni Kumar, P. 1994 ApJ, 428, 827

\refind  Kumar, P, Fardal, M.A., Jefferies, S.M., Duvall, T.L. Jr., 
Harvey J.W., and Pomerantz, M.A. 1994, ApJ, 422, L29

\ni Roxbourgh, I.W. \& Vorontsov, S.V., 1995, MNRAS, 272, 850

\vfill\eject

\centerline{\bf FIGURE CAPTIONS}

\ni{\bf FIG 1} Two peaks with values of the asymmetry parameter $\eta_\alpha$ of -5\% and -10\%; see the text for the definition of $\eta_\alpha$.  Both spectra were calculated using the 1-D model with $a=3275$ seconds of sound travel time and $\nu=1.5$ mHz.  For the peak with 
$\eta_\alpha=-10\%$, the source was placed at the upper turning point; for the other, the source was 
placed 300 km below the upper turning point.  Line profiles with the same value of the asymmetry 
parameter $\eta_\alpha$ look essentially identical, regardless of source location, frequency, cavity 
length, linewidth, or type of damping.

\ni{\bf FIG 2} a) Asymmetry in the 1-D model (quantified in terms of the parameter $\eta_\alpha$) 
as a function of source height above the upper turning point (h) 
for modes in cavities of approximate length 3275s, 1625s, and 500s.  
(Cavity lengths are measured in terms of sound travel time.)  
Points where $\eta_\alpha$ abruptly changes from positive to negative (with decreasing h) 
correspond to nodes of the respective eigenfunctions.  
Linewidth was fixed at 5 $\mu$Hz for all of these calculations.  
b) Also shown is asymmetry in the 1-D model as a function of linewidth for modes 
in cavities of length 3275s and 475s.  Mode frequency was fixed at 3.0 mHz for these calculations.

\ni{\bf FIG 3}  Plotted on the same set of axes are: i) The ratio of the amplitude of the cavity wave 
to the amplitude of the direct wave (this ratio varies symmetrically about its peak), 
ii) The phase of the cavity wave with respect to the direct wave (in radians), and 
iii) The ratio of the amplitude of the reflected wave to 
the amplitude of the direct wave.  
(All three curves were calculated using the 1-D model, with the source outside the cavity, 
for a peak with $\eta_\alpha = -10\%$.)

\ni{\bf FIG 4} Asymmetry in the solar model (quantified in terms of the parameter $\eta_\alpha$) 
as a function of source depth (measured relative to the top of the convection zone) 
for modes with $\ell \approx$ 5, 100, 250, and 375.  
Points where $\eta_\alpha$ changes abruptly from positive to negative 
(with increasing depth) correspond to nodes of the respective eigenfunctions.

\ni{\bf FIG 5} Frequency shifts (expressed as percentages of corresponding linewidths) of best-fit 
frequencies from eigenfrequencies, versus percentage asymmetry $\eta_\alpha$.  These points 
correspond to the peaks, calculated using the solar model, whose asymmetries are plotted in Figure 4.  
The slope of the line is approximately 1.5.

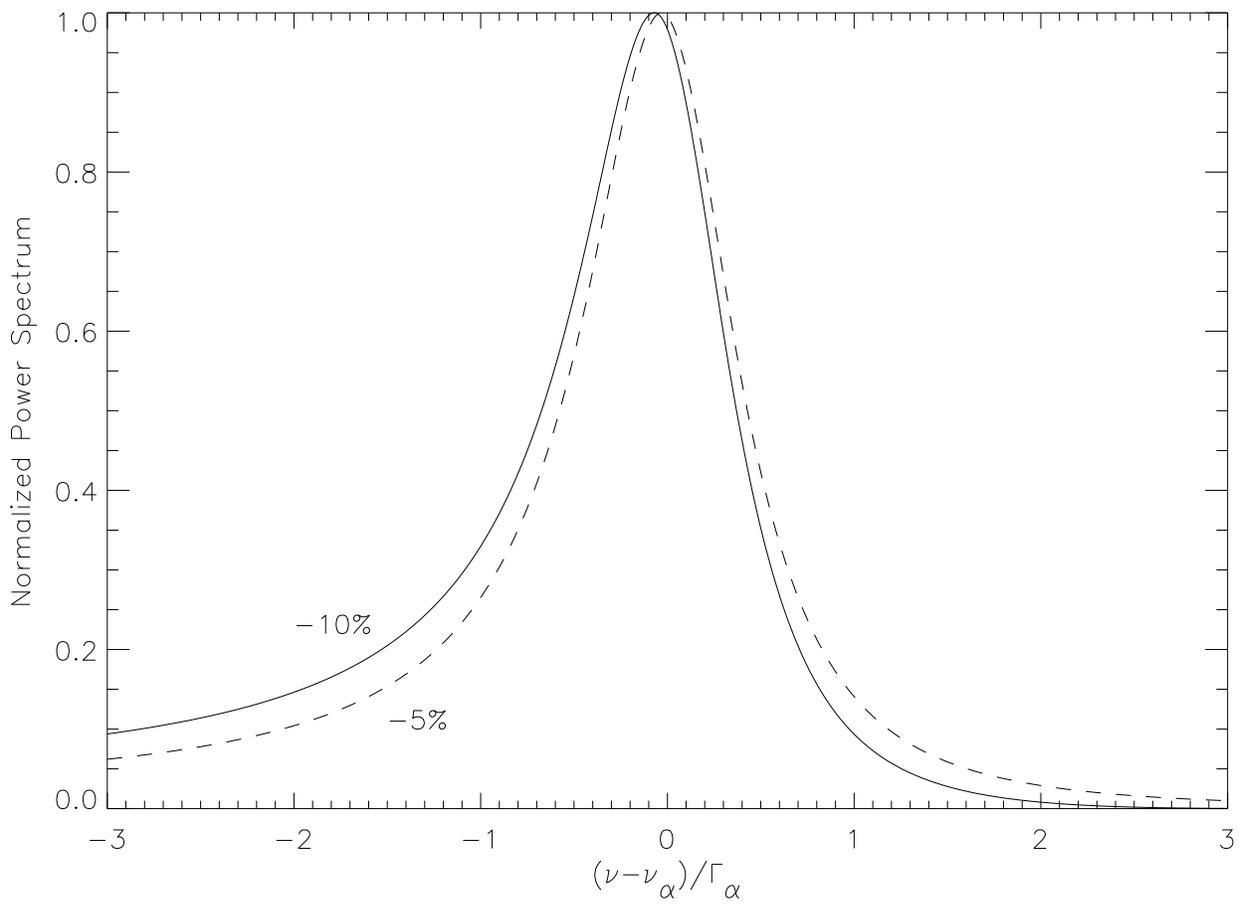
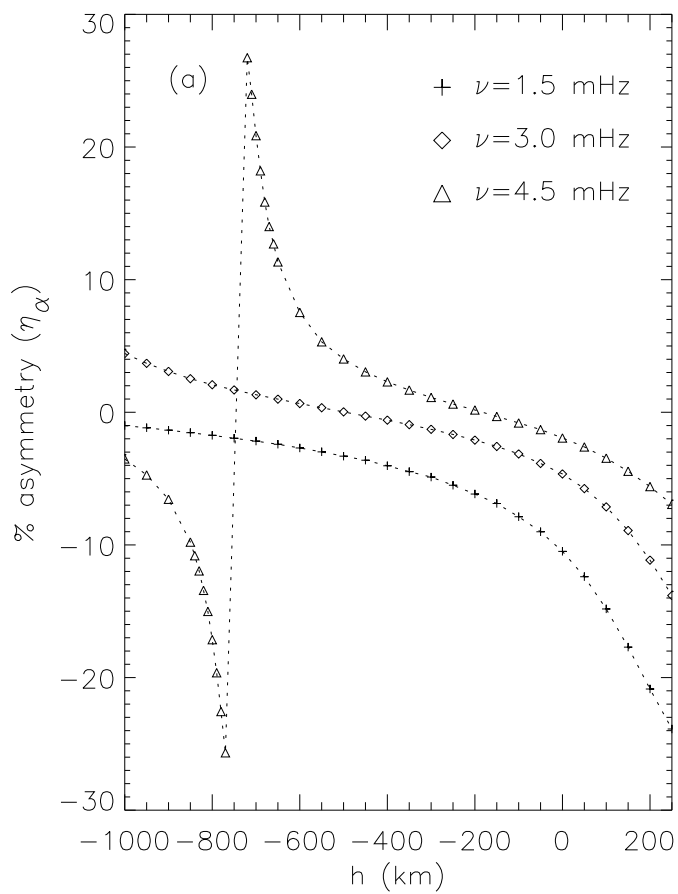
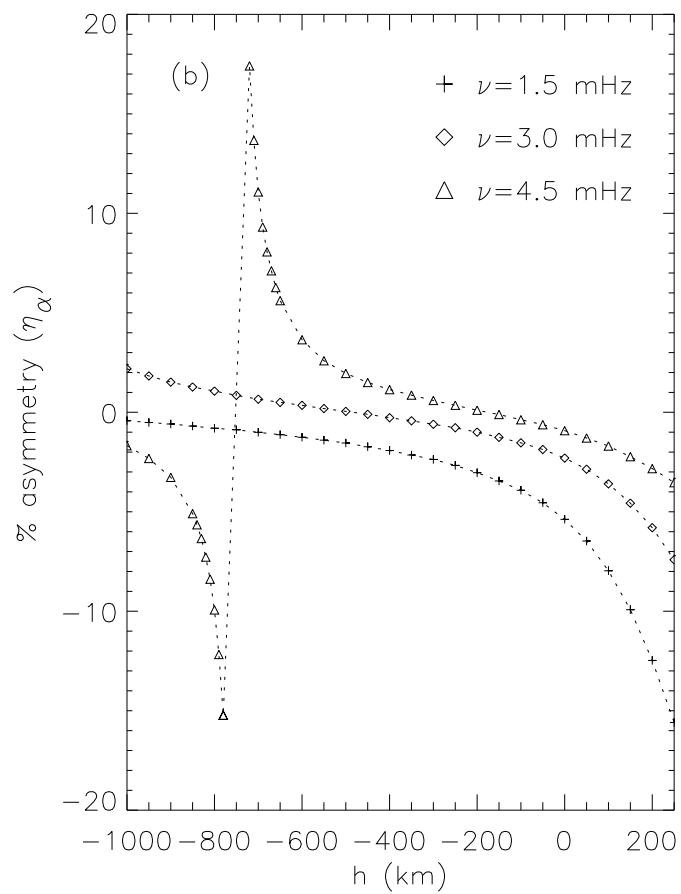
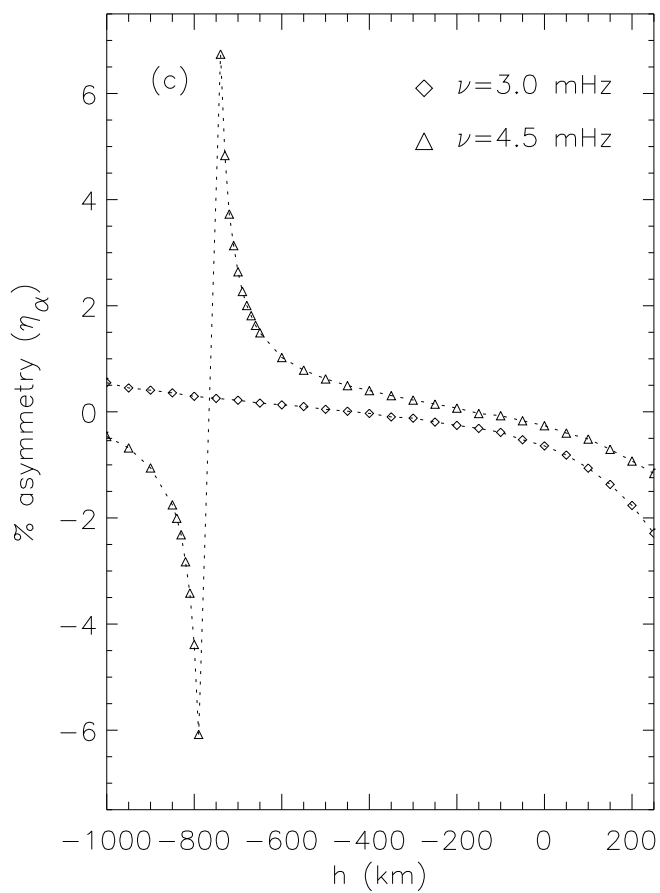
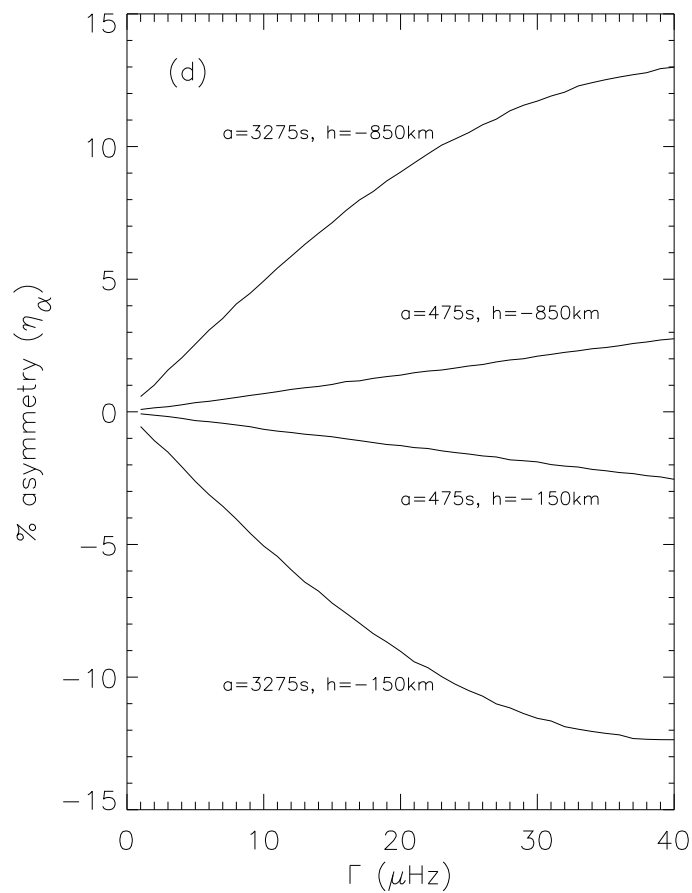
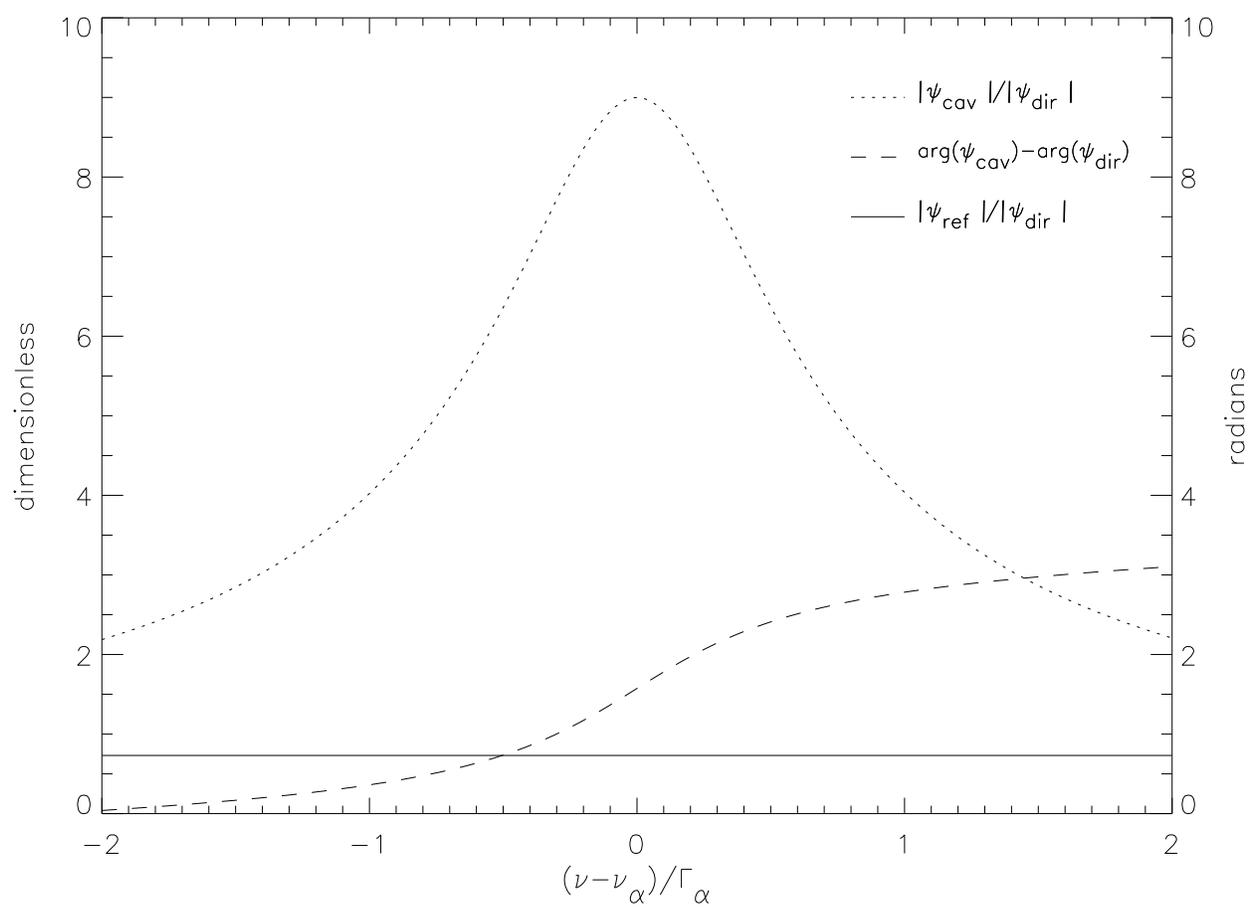
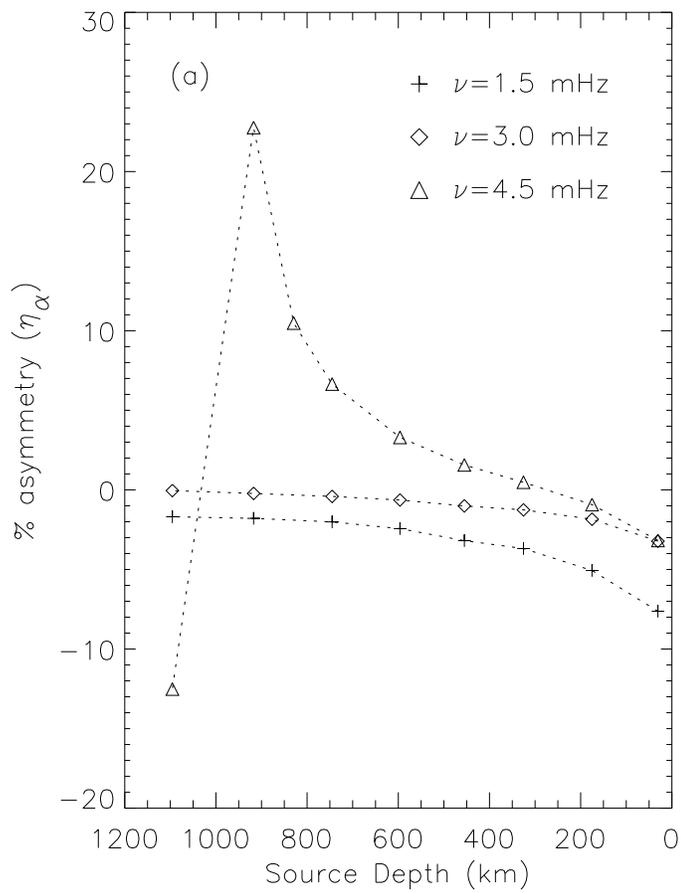
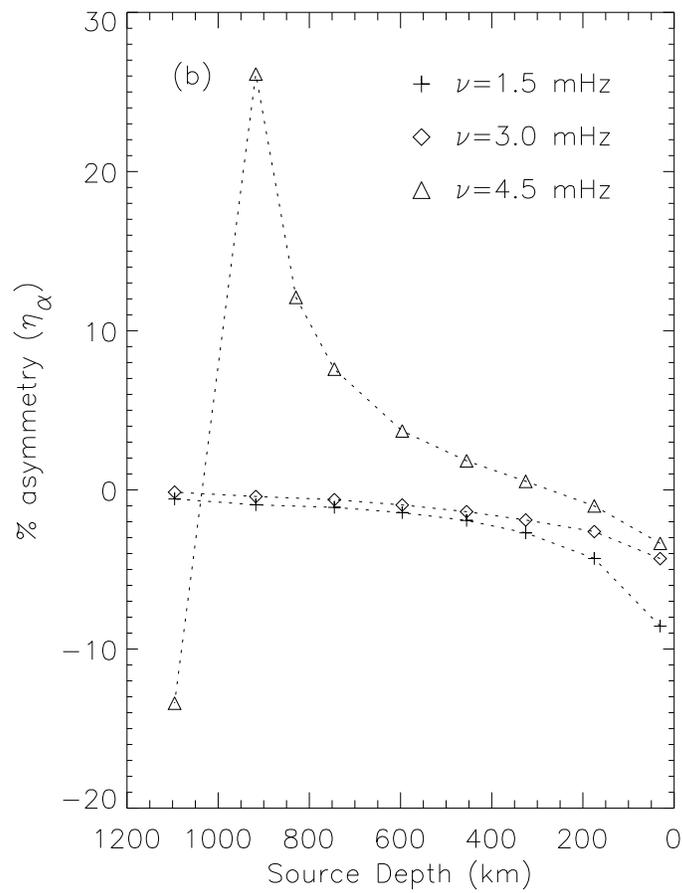
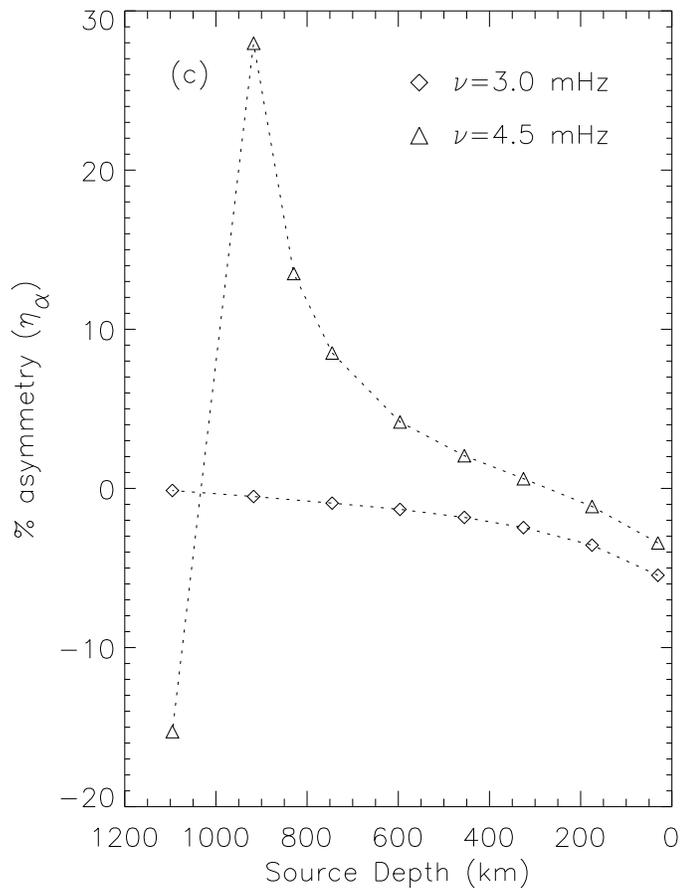
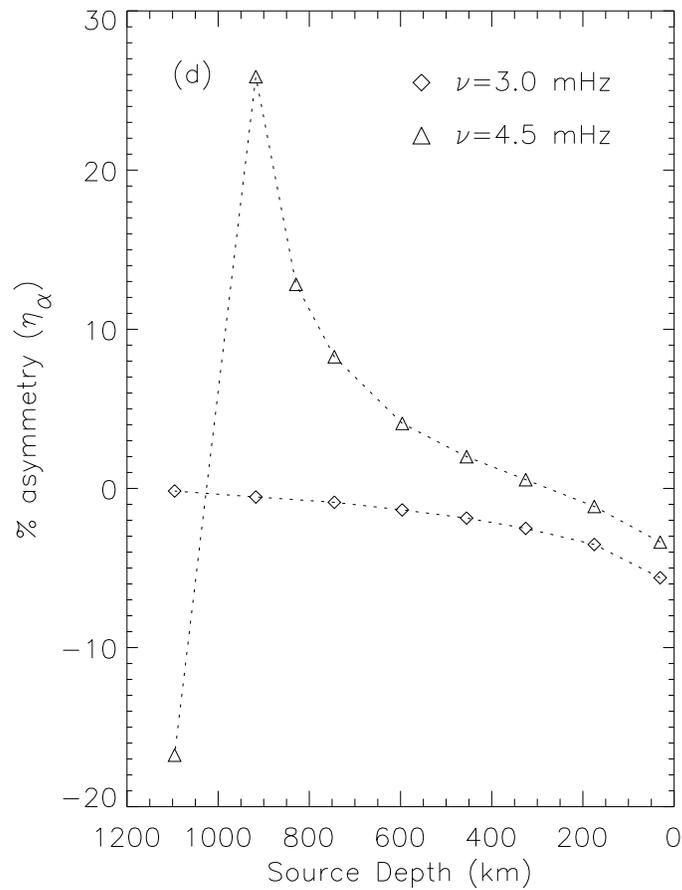
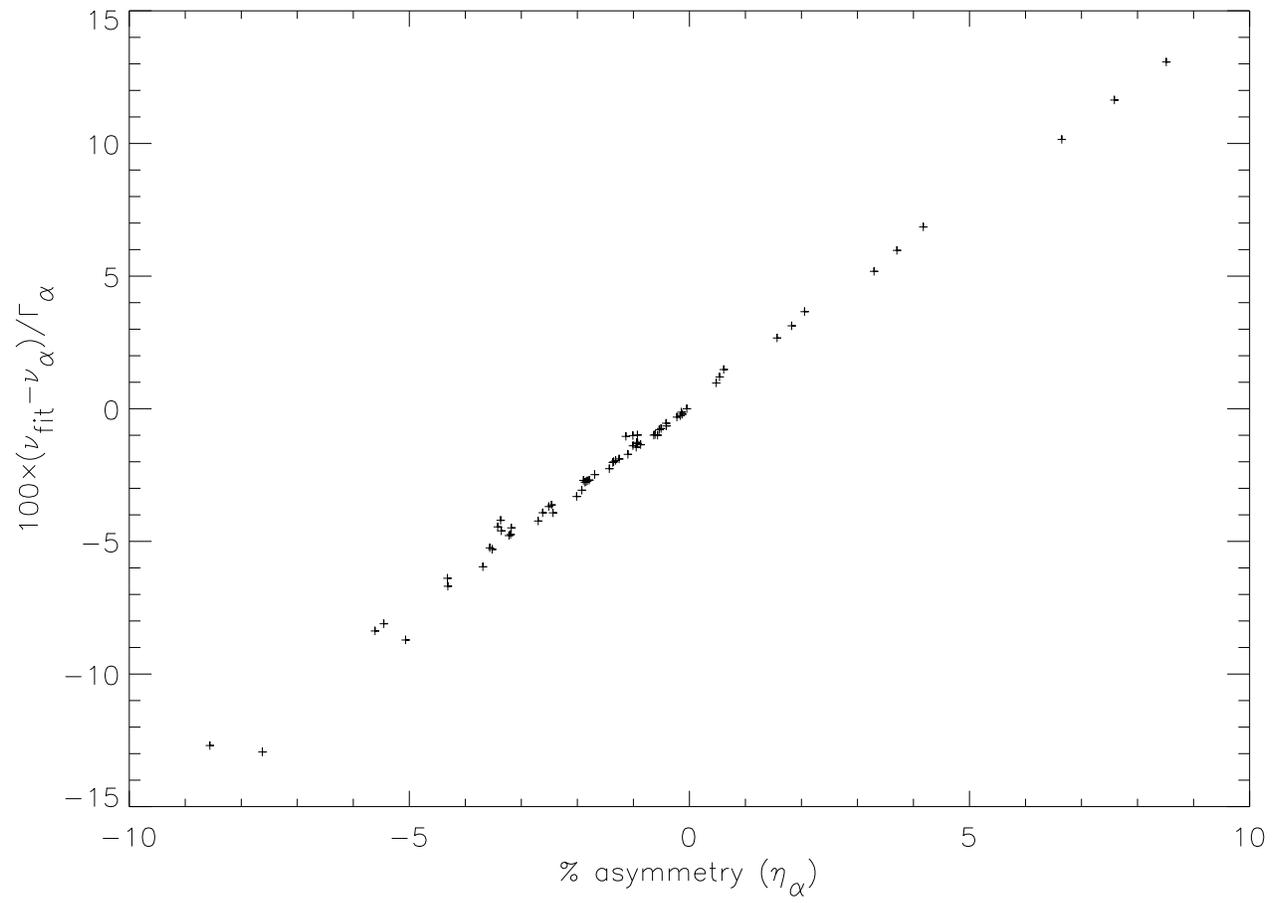
\bye